# First demonstration of temperature control enabled high power mode-switchable fiber laser

Jiaxin Song，Haiyang Xu，Hanshuo Wu，Liangjin Huang, Jiangming Xu, Pu Zhou*
College of Advanced Interdisciplinary Studies, National University of Defense Technology, Changsha, China
*zhoupu203@163.com

**Abstract**—A transverse mode-switching method was proposed and demonstrated in a high-power ytterbium-doped fiber oscillator. 17.8 W $LP_{11}$ mode laser was obtained, and it could be switched to 16.5 W $LP_{01}$ mode laser through temperature control.

**Keywords**—transverse mode switchable, Yb-doped fiber laser, temperature control, high power

## I. Introduction

Fiber lasers with specific higher-order modes are desired in many applications owing to their unique spatial intensity, phase and polarization distributions [1-5]. Therefore, fiber lasers with transverse mode switchable property are desired because of flexibility and universal applicability. Traditional methods of implementing mode switchable operation includes using few-mode fiber Bragg grating (FBG) combined with a tunable filter [6], a spatial light modulator [7], or a polarization controller [8], active control using algorithm [9] or acousto-optic modulation [10], and applying pressure to long-period grating (LPG) [11] etc. Through power amplification, several hundred-watt level transverse mode-switchable fiber lasers have been obtained both in continue-wave and pulsed operations [9, 11]. However, the power of the mode-switchable fiber oscillator is almost no more than 1 W [6, 12-13]. Further power scaling of mode-switchable fiber laser could reduce the amplification stages and make the whole system more compact.

In this paper, we report a mode switchable Yb-doped fiber oscillator through temperature control of few-mode FBGs. Pure LP11 mode fiber laser based on few-mode FBGs has been obtained in our previous experiment [14]. By tuning the temperature, LP01 mode laser could also generate in the same cavity. Around 17 W output power was obtained and almost kept unchanged during mode switching process, which, so far as we know, is the record power in terms of transverse mode-switchable all-fiber laser.

## II. Experimental Setup

The experimental setup is shown in Fig. 1. Two laser diodes (LDs) were connected with the input ports of the combiner to serve as pump source. A piece of 3.5-meter-long Yb-doped fiber (YDF) was used as gain medium, the core and cladding diameters of which are 15 and 130 μm, respectively. Cladding mode stripper, which is made of a high refractive index glue, was coated on the bare fiber around the splicing point to remove the cladding light and protect the high reflectivity (HR) FBG. Both the two output ends are angle-cleaved with 8 ° to avoid unexpected reflection.

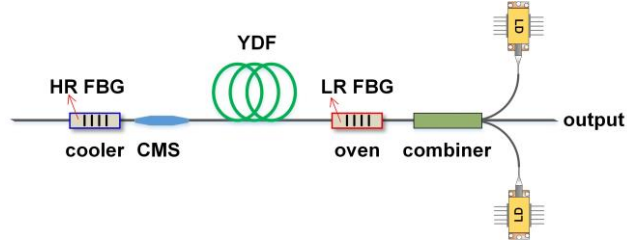

Fig. 1. Experimental setup of mode switchable Yb-doped fiber laser. HR: high reflectivity, FBG: fiber Bragg grating, CMS: cladding mode stripper, YDF: Yb-doped fiber, LD: laser diode.

The HR FBG was written on a piece of SMF-28 fiber, the reflectivity of which is 98.1 %. The core and cladding diameters of SMF-28 fiber is 8.2 and 125 μm, respectively. In contrast, a piece of 15/130 μm was selected to fabricate the low reflectivity (LR) FBG. The reflectivity of the LR FBG is 73.7 %, which is higher than that in traditional fiber laser configuration in order to enhance the gain. Both the SMF-28 fiber and 15/130 fiber are few mode fibers and support $LP_{01}$ and $LP_{11}$ mode [15]. By using FBGs with different types of fiber, the reflection peaks corresponding to $LP_{11}$ mode could match, while that of $LP_{01}$ mode could be staggered as shown in Fig. 2. As the ambient temperature of the FBG increases, the reflection peaks would show redshift trend [15]. In this experiment, the HR FBG was placed on a cooler, while the LR FBG was put on an oven. The temperature difference between the cooler and over could reach 60 K. In this way, the reflection peaks of the $LP_{01}$ modes could be matched by adjusting the temperature difference. It is the basic mechanism of mode switchable operation. Considering the mismatch of core diameters between the YDF and HR FBG, backward pump configuration was chosen.

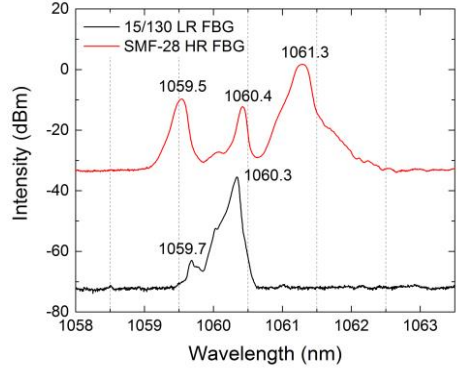

Fig. 2. The reflection spectra of the HR FBG and LR FBG

III. RESULTS AND DISCUSSION

At room temperature, i.e. around 298 K, the $LP_{11}$ mode laser would lase. As the pump power increases, the FBGs would be heated up. In this case, the temperature control platform only plays a role of maintaining stability. The central wavelength of $LP_{11}$ mode laser at the pump power of 3.5 W is 1059.4 nm, as shown at the bottom of Fig. 3. As the temperature rises, another wavelength component at 1060.1 nm, i.e. $LP_{01}$ mode, occurs as shown in the middle of Fig. 3. The intensity of $LP_{01}$ mode component gradually exceeds that of the $LP_{11}$ mode. According to the spectra evolution in the experiment, the central wavelength of output laser would redshift about 0.02 nm when the temperature difference increases by 1 K. Since the interval between the reflection peaks of $LP_{01}$ mode and $LP_{11}$ mode is ~0.9 nm, it is estimated that about 45 K temperature difference is required to realize mode switching process. In the experiment, the temperature difference continues to be increased. When the temperature difference becomes 49 K, pure $LP_{01}$ mode is obtained as depicted at the top of Fig. 3. The corresponding beam profiles were recorded by a CCD camera and presented as insets. Fig. 4 gives the output power curve at these two temperature as the pump power grows. The maximum power of $LP_{11}$ mode output is 17.8 W with the slope efficiency of 44.4 %. The maximum output power of $LP_{01}$ is 16.5 W, and the slope efficiency is 43.3%. By comparison, the output power and efficiency of the two operation modes are nearly the same. The results above verifies that the mode-switching method based on temperature control of few-mode FBGs could be applied in high power fiber laser.

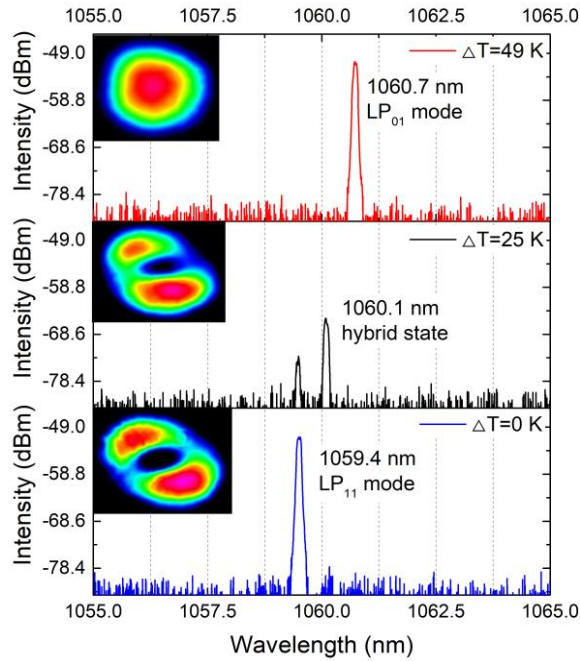

Fig. 3. The evolution of the output spectra from $LP_{11}$ mode (bottom) to hybrid state (middle) to $LP_{01}$ mode (top); and the corresponding beam profile at the pump power of 3.5 W (inset)

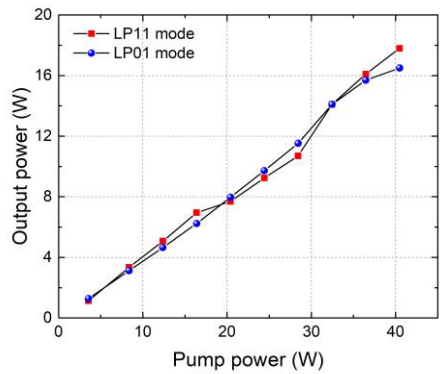

Fig. 4. The power curve of the $LP_{11}$ and $LP_{01}$ mode fiber laser

IV. CONCLUSION

In conclusion, we demonstrated a mode switchable Yb-doped fiber laser by adjusting the temperature of few-mode FBGs. 17.8 W $LP_{11}$ mode fiber laser was obtained with the slope efficiency of 44.4 %, while the output power of $LP_{01}$ mode fiber laser could reach 16.5 W with the slope efficiency of 43.3 %. To the best of our knowledge, it is the first demonstration of mode-switchable fiber laser, and the output power is increased by 2 orders of magnitude compared with previous demonstrations. The results show that temperature controlling method could provide a practical laser source for applications that require mode switching and high power operation.

## Acknowledgment

The authors thank Prof. Zefeng Wang, Dr. Meng Wang and Miss Xiaoyu Hu for providing the FBGs, and also thank Mr Jun Ye for his valuable advice.

## References

[1]. K. Venkatakrishnan, "Generation of Radially Polarized Beam for Laser Micromachining," Journal of Laser Micro, vol. 7, pp. 274-278, Nov 2012.

[2]. S. Ramachandran, J. W. Nicholson, S. Ghalmi, M. F. Yan, P. Wisk, E. Monberg, and F. V. Dimarcello, "Light propagation with ultralarge modal areas in optical fibers," OPT LETT, vol. 31, pp. 1797, June 2006.

[3]. Y. Xiong, R. B. Priti, O. Liboiron-Ladouceur, Y. Xiong, R. B. Priti, O. Liboiron-Ladouceur, Y. Xiong, R. B. Priti, and O. Liboiron-Ladouceur, "High-speed two-mode switch for mode-division multiplexing optical networks," OPTICA, vol. 4, pp. 1098-1102, September 2017.

[4]. B. Sun, A. Wang, L. Xu, C. Gu, Z. Lin, H. Ming, and Q. Zhan, "Low-threshold single-wavelength all-fiber laser generating cylindrical vector beams using a few-mode fiber Bragg grating," OPT LETT, vol. 37, pp. 464-466, February 2012.

[5]. C. Varin, and M. Piché, "Acceleration of ultra-relativistic electrons using high-intensity $TM_{01}$ laser beams," Appl. Phys. B, vol. 74, pp. s83-s88, June 2002.

[6]. B. Sun, A. Wang, L. Xu, C. Gu, Y. Zhou, Z. Lin, H. Ming, and Q. Zhan, "Transverse mode switchable fiber laser through wavelength tuning," OPT LETT, vol. 38, pp. 667-669, March 2013.

[7]. A. Forbes, A. Dudley, and M McLaren, "Creation and detection of optical modes with spatial light modulators," ADV OPT PHOTONICS, vol. 8, pp. 200, April 2016.

[8]. X. Du, H. Zhang, P. Ma, X. Wang, P. Zhou, and Z. Liu, "Spatial mode switchable fiber laser based on FM-FBG and random distributed feedback," LASER PHYS, vol. 25, pp. 95102, August 2015.

[9]. R. Su, B. Yang, X. Xi, P. Zhou, X. Wang, Y. Ma, X. Xu, and J. Chen, "500 W level MOPA laser with switchable output modes based on active control," OPT EXPRESS, vol. 25, pp. 23275, September 2017.

[10]. J. M. Daniel, and W. A. Clarkson, "Rapid, electronically controllable transverse mode selection in a multimode fiber laser," OPT EXPRESS, vol. 21, pp. 29442, November 2013.

[11]. T. Liu, S. Chen, X. Qi, and J. Hou, "High-power transverse-mode-switchable all-fiber picosecond MOPA," OPT EXPRESS, vol. 24, pp. 27821, November 2016.

[12]. Y. Shen, G. Ren, Y. Yang, S. Yao, Y. Wu, Y. Jiang, Y. Xu, W. Jin, B. Zhu, and S. Jian, "Switchable narrow linewidth fiber laser with $LP_{11}$ transverse mode output," OPT LASER TECHNOL, vol. 98, pp. 1-6, January 2018.

[13]. L. Carrion-Higueras, E. P. Alcusa-Saez, A. Diez, and M. V. Andres, "All-fiber laser with intracavity acousto-optic dynamic mode converter for efficient generation of radially polarized cylindrical vector beams," IEEE PHOTONICS J, vol. 1, pp. 1500501, February 2017.

[14]. J. Song, H. Xu, H. Wu, L. Huang, J. Xu, H. Zhang, and P. Zhou, "High power narrow linewidth LP11 mode fiber laser using mode-selective FBGs," LASER PHYS LETT, vol. 15, pp. 1-5, September 2018.

[15]. T. Mizunami, T. V. Djambova, T. Niiho, and S. Gupta, "Bragg gratings in multimode and few-mode optical fibers," J LIGHTWAVE TECHNOL, vol. 18, pp. 230-235, February 2000.